# Introducing Political Ecology of Creative-Ai

Andre Holzapfel[1] (https://orcid.org/0000-0003-1679-6018)

[1] Division of Media Technology and Interaction Design, KTH Royal Institute of Technology

**Abstract (150 words)**

This chapter introduces the perspective of political ecology to the application of artificial intelligence to artistic processes (Creative-Ai). Hence, the environmental and social impact of the development and employment of Creative-Ai are the focus of this text, when we consider them as part of an economic system that transforms artistic creation to a commodity. I first analyse specific Creative-Ai cases, and then conduct a speculation that takes Jacques Attali's writing on the role of music in society as a vantage point, and investigates the environmental and social consequences of an automatic composition network controlled by a large music streaming platform. Whereas the possibilities that emerge from Creative-Ai may be promising from an artistic perspective, its entanglement with corporate interest raises severe concerns. These concerns can only be addressed by a wide cross-sectoral alliance between research and arts that develops a critical perspective on the future directions of Creative-Ai.

Keywords (6 keywords): political ecology, artificial intelligence, arts, computational creativity, sustainability, ethics

## Introduction

Applications of artificial intelligence to artistic processes (Creative-Ai[1]) are becoming more common, and media attention towards these applications is large. At the focus of many media contributions is the question when a human artist-genius will be replaced by a similarly genius and autonomous Ai. But a quick look at various widely discussed Ai art projects clarifies that in most cases the Ai acts far from autonomously, and that the symbolic realm of the artistic creation is densely connected to a material dimension. The immersive media installation titled "Archive Dreaming" by Refik Anadolu creates a "canvas with light and data applied as materials", and facilitates both user-driven and automatic exploration "of unexpected correlations among documents". The project was realised with the support of 20 collaborators at three institutions, and the Ai comprises a model that has been trained on 1.7 million documents using the computational power provided to the artist by Google's Artists and Machine Intelligence Program (Anadol 2017). The completion of the composition of Beethoven's 10th symphony was approached by a group of music historians, musicologists, composers and computer scientists over a period of several years involving extended data preparation and Ai model design (Elgammal 2021). The novel "1 the Road" was written by an Ai, but the artist Ross Goodwin carefully orchestrated the sensor inputs that would feed the Ai model during a road trip. The creative process involved not only Goodwin, his car, and an Ai, but also a film crew accompanying him and documenting the road trip from New York to New Orleans (Hornigold 2018). In comparison to the Ai Beethoven project, however, the artist took the deliberate choice to present the Ai output in its raw form, with the motivation to illustrate limitations and potential of such an autonomous machine creation.

---
[1] The reference to artificial intelligence as Ai with lower-case i, is a conscious choice that reflects the current state of Ai systems as not (yet) fulfilling plausible minimum criteria of intelligence.

Besides the formation of an Ai model in the context of a specific artwork, many companies have shaped tools for media production that use - or at least claim the use - Ai as a means to support the creative processes of the users of the tools. Examples are the use of Ai for mastering the sound of a music production[2], audio restoration[3], or the scene analysis and filtering of images based on Ai by Photoshop (Clarke 2020). Such systems have been called creative support tools (CST), and these tools have been well-documented within HCI research for over a decade (Frich et al. 2019; Gabriel et al. 2016). Based on the literature review of 143 papers, Frich et al. (2019) define CST as a tool that "runs on one or more digital systems, encompasses one or more creativity-focused features, and is employed to positively influence users of varying expertise in one or more distinct phases of the creative process". Following this definition, a Creative-Ai tool is simply a CST that employs artificial intelligence. Demystifying this latter buzzword, artificial intelligence in the context of Creative-Ai (and in most other contexts) is nothing but data-driven analysis and/or synthesis of media data, and the intelligence that emerges from training processes can be fairly considered as narrow. That is, all tools for artists and designers so far comprise an intelligence that can at least partially conduct a creative process in a fairly restricted context. The Beethoven Ai can continue phrases in the style of one specific composer, and the Archive Dreamer can hallucinate novel data in the context of one specific archive. Going beyond these borders demands adapting and re-training models, which demands on the material side new collections of data and expenditure of energy for the training of models.

The goal of this chapter is to develop a perspective on the environmental impact of creative processes that make use of Ai, when we consider them as part of an economic system that transforms artistic creation to a commodity. I will discuss specific Creative-Ai cases and their environmental impact along with the role of large corporations involved in Creative-Ai development. Already in the 1970s, the economic and social theorist Jacques Attali (1985) predicted a situation that strongly relates to current business plans by the largest music streaming provider Spotify. As we will see, prospects of using Creative-Ai as a collaborative tool that supports creative processes promise an extension of the horizon of artistic possibilities, but such optimistic perspectives neglect the role of large corporations as gatekeepers to systems and infrastructures. I believe that it is timely to combine political, environmental, and sociocultural perspectives on Creative-Ai to better understand how the power exercised by capital and the environmental impact of Creative-Ai may motivate researchers and artists to adopt a more critical perspective. The chapter - written by an ethnomusicologist/computer scientist - is characterised by a certain focus on music, but its general implications can be transferred to other forms of art.

---

[2] https://www.landr.com/
[3] https://www.izotope.com/en/products/rx.html

## Related Work

The application and development of Creative-Ai has been investigated from various perspectives. A central venue of discussing Creative-Ai is the International Conference on Computational Creativity (ICCC), at which practical Ai applications in various art forms (e.g. Wertz and Kuhn 2022; Hertzmann 2022; Tan 2019) are discussed as well as sociocultural and theoretical dimensions of applying computational tools in creative processes ( e.g. Riccio et al 2022; Brown et al 2021). Within the Conference on Human Factors in Computing Systems (CHI) the interfaces and interactions with Creative-Ai have been explored (Oh et al 2018; Ragot et al 2020), and in 2022 a dedicated workshop gathered researchers working with Ai in the context of generating art (Muller et al 2022). More specialised HCI related venues, such as the International Conference on New Interfaces for Musical Expression (NIME) feature to an increasing amount discussions of how Ai can present novel and inspiring ways to interact with music technology (Tahiroğlu et al 2020; Murray-Browne and Tigas 2021). In addition, many machine learning conferences feature special sessions that focus on novel model architectures in the context of artistic creation[4].

Social and ethical implications of Creative-Ai related to the use of training data (Sturm et al 2019), and impact on diversity (Porcaro et al 2021) and fairness (Ferraro et al 2021) in diverse cultural contexts have been discussed to some extent. When it comes to potential environmental consequences of Creative-Ai use, it is unclear how large the energy consumption and related carbon footprints of specific artistic projects may be (Jääskeläinen et al. 2022). Whereas the environmental impact of individual artistic projects may be negligible, the environmental perspective becomes relevant when Creative-Ai tools are becoming a commodity to which access to larger user groups is provided by a corporation. This nexus of environment, society, and politics is what is at the core of studies of political ecology (Biersack 2006). As Watts (2017) states, political ecology studies "forms of access and control over resources and their implications for environmental health and sustainable livelihoods". The severity of such implications in the context of Ai and Big Data has been analysed, for instance, by Crawford (2021), Velkova (2016) and Thylstrupp (2019). A ground-breaking study of the political ecology of the music industry has been conducted by Devine (2019), who demonstrates that the digitalisation of the music industry in the age of streaming increased rather than decreased its carbon footprint, compared to the era of physical distribution of music media. In this rather grim picture of the music industry, Devine has not yet taken into account the use of Ai in the context of recommendation and content generation.

A series of publications has directed attention to the energy consumption of the use of Ai in various fields. One of the most widely perceived publications (Strubell et al., 2019) estimates the energy consumption related to the development of a deep learning based model in the context of natural language processing (NLP). Strubell et al. (2019) reveal that their development of an NLP model within six months involved the use of 27 years of GPU-time[5] for training models, with massive financial costs and energy consumption for cloud computing. Besides the environmental impact, the authors emphasise that building such models becomes a privilege of well-funded academic institutions and corporations. Therefore various authors suggest documenting energy and carbon footprints of model development and training along common performance metrics (Strubell et al 2019; Anthony et al 2020). Tools for estimating environmental impact are increasingly available (Anthony et al 2020; Henderson et al 2020), and although the estimates may have a large error margin they are helpful to provide an orientation of the order of $CO_2$ emissions caused by the development.

---

[4] For instance, since 2017 the Machine Learning for Creativity and Design workshop in the context of the NiPS conference (https://nips2017creativity.github.io/), and the special track on Ai, the Arts, and Creativity at the IJCAI-ECAI 2022 conference (https://ijcai-22.org/calls-creativity/).

[5] Most deep learning development makes use of graphics processing units (GPUs), instead of the central processing unit (CPU) of a computer. Using several years of computing time in a shorter time span becomes possible by using a larger number of GPUs in parallel, usually at academic or commercial data centres. This way, 27 years of single GPU time may be spent by, for instance, 1500 GPUs continuously running in parallel for less than a week.

## Environmental impact of creative-Ai

In a recent publication (Jääskeläinen et al. 2022), we mapped the aspects that need to be taken into account when estimating the environmental impact of a specific artistic process that involves the use of creative-Ai. We identified practice-related aspects, such as running an algorithm in iterations of slightly different constellations, and the varying amount of such energy-intense steps in the various phases of an artistic process. As of now, our knowledge of how these practice-related aspects manifest themselves within specific artistic processes is very limited. Our own research aims to utilise diary studies with artists with the goal to close this knowledge gap, as the information provided in relation to already published artworks remains too sparse when it comes to the details of the artistic process. As the second group, we identified material aspects of creative-Ai, such as the employed hardware, the type of energy source, model architecture, and training data. On this side, information from existing Ai-art projects allows us to obtain an overview of most common architectures and training data sizes. On these grounds, the hypothesis is that the environmental impact of creative-Ai use may be quite significant, since the practice-related aspects involve repeated training of various formations of a model using various datasets, whereas the material aspects involve the use of model architectures that comprise a large number of parameters that need to be optimised using specialised GPU hardware.

It is essential for an understanding of the political ecology of creative-Ai to indicate the orders of energy consumption and/or carbon emissions related to training and developing certain models used in artistic contexts. As with all machine learning models, there is the training phase in which the model parameters are determined based on the training data, and the inference phase, in which a trained model is employed to produce an output. We will focus our considerations on the training phases, as the inference is typically - in relation to the training - less energy-intense. The model GPT-3[6] is a very large NLP model that produces text outputs based on prompts, and it has been used, for instance, to generate poetry[7]. Based on the calculations by Anthony et al (2020) the training of that model took about as much energy as it would take to drive around the globe by car almost 18 times.[8] Setting this in relation to per capita consumption of energy in Sweden (Enerdata 2020) - the country with the second-highest per per capita consumption in Europe - this is about the yearly per capita consumption of 15 Swedish citizens.  As OpenAi does not provide information of how much energy the whole R&D process required, let us assume in analogy to Strubell et al (2019) that the energy consumption of the overall R&D may have been three orders larger than a training the network a single time, which would send us around the planet 18,000 times by car, or supply a small Swedish city with energy for one year. The list of applications that employ GPT-3 in their functionalities[9] consists to about 15% (57 out of 353) of applications related to creative processes in arts and the creative industries, which would mean that almost 3000 of our planet round trips are attributed to Creative-Ai applications.

---

[6] https://openai.com/api/
[7] See, for instance, Branwen (2022) for a detailed documentation of such artistic processes and outcomes.
[8] Without intention of endorsing excessive car driving and its environmental impact, I will use this comparison throughout the text because I hope it works as a vivid illustration for most readers.

But let us turn our attention to individual artistic projects instead of looking at a specific model. Throughout the last few years, several large IT companies have established funding schemes for artist residencies[10]. Within the NVidia artist gallery, we encounter the work of various artists who have used NVidia's infrastructures to realise artistic projects. The project "Strange Genders" by 64/1 and Harshit Agrawal employs an NVidia StyleGAN2, which the artists trained on a custom dataset of hand-drawn figures. According to the original paper (Karras et al 2020) the energy required to train this model with its 69 million parameters from scratch is 0.68MWh. Using the carbon intensity of the US in the year the artwork was produced as a basis (0.39 kgCO2/kWh, according to EIA (2021), training the model once would have caused about 265 kgCO2eq, which would take us about 2200 km far with an average car (following the calculation method by Anthony et al (2020)).
Japan-based artists Nao Tokui and Qosmo employed a variety of deep learning models in their project Neural Beatbox/Synthetic Soundscape. They combine the use of Variational Autoencoders for generating rhythms, Convolutional Neural Networks for sound classification, Recurrent Neural Networks (SampleRNN) for soundscape generation, and GAN to obtain original drum patterns. It is impossible to arrive at estimates of the energy invested for the overall combination of models, as the energy consumption for training them has not been specified in most of the original papers. However, the training of the SampleRNN is referred to have taken "a few days" despite the availability of NVidia resources, which is consistent with the training time of one week on a single GPU as specified by Mehri et al (2017). Using again the carbon intensity of 0.39 kgCO2/kWh, 250W as the maximum power specified for the chip by NVidia , and a power usage effectiveness (PUE) of 1.125 as by Anthony et al (2020), we arrive at a energy consumption of 168h * 0.25kW * 1.125 = 47.25kWh and a related carbon emission of 18.43kgCO2eq. While this is only a short car ride (about 150km), the estimate is very optimistic as the PUE of a single personal computer with a GPU is larger than that of a highly optimised data centre.

These two examples indicate that artistic projects that involve the use of Creative-Ai may involve large energy consumption for training the needed networks. While we do not know details of the artistic processes, it is likely that such processes involve a similar amount of re-iteration and refinement as reported for engineering projects by Strubell et al (2019). Whereas engineering projects optimise towards a limited amount of usually quantitative objectives, the artistic process is much more open-ended and involves qualitative criteria imposed by the artist. As there is no reason to assume that this process is simpler, the development of an Ai for artistic purposes can involve many iterations of training and result in energy consumption that is orders higher than training a network a single time.

---

[9] https://gpt3demo.com/
[10] Examples are Google (https://ami.withgoogle.com/) and NVidia (https://www.nvidia.com/en-us/research/ai-art-gallery/).

## Political Ecology of Creative-Ai

All this evidence implies that the carbon footprint resulting from artistic projects involving Creative-Ai is likely to be large compared to artistic projects with similar outcomes but not involving Creative-Ai. More important than raising awareness of this energy consumption is to discuss its consequences for artist populations on a larger scale, which is where the frame of political ecology becomes relevant. Access to large computing power is not available to all (Strubell et al 2019), and it is focused on institutes and corporations in the Global North. Many powerful Ai-models used in creative processes are trained models that can be accessed through the APIs provided by the developing companies, and the availability depends on the goodwill and the business plans of the companies. If individual artists wish to compute outputs from such models, or even wish to adapt or train new models to their own data, then they are depending on large IT companies providing access to resources.

An equitable engagement of all stakeholders of Ai applications into a discussion of the environmental impact of Ai has been argued to be a cornerstone of a third wave of Ai ethics (van Wynsberghe, 2021), with the possible conclusion to refrain from using Ai in certain application areas. Rephrasing van Wynsberghe, one has to ask if algorithms that can paint or compose are really worth the environmental cost. While there is no doubt that other industry sectors have a larger carbon footprint than the creative industries, it is nevertheless an ethical question for the artists if they want to employ a technology that dramatically increases the environmental impact of their work, and that puts their ability to practise art and secure their livelihood at the goodwill of IT corporations. However, the environmental impact is in most cases not obvious for the artist. As in the case of Nao Tokui and Qosmo, this impact presents itself in the form of longer processing time without any further information of the actual energy consumption during this time. With most artists who employ Creative-Ai having acquired the programming skills that are needed to run inference and training of models in environments such as python, it seems a realistic suggestion to employ tools to estimate energy consumption of the development and to report such estimates when publishing the artwork.

As we observed in Jääskeläinen et al (2022b), in the design of interactions and experiences with Creative-Ai all current interfaces to Ai-models have distanced the actual consequences of the actions very far away from the users of these technologies: the amount of energy required to perform some inference and/or training remains concealed, just as the kind of energy that was employed by some remote servers involved in the computations. We argue that one consequence of current interface design is that it may promote slow violence (Nixon 2011), i.e. a lack of transparency towards environmental consequences that develop on another temporal and geographical scale than the artistic process. Therefore, we consider it important to design interfaces for Creative-Ai that stimulate the development of care to the ecosystem, a form of care that we propose to call "slow care". Such design is situated in the framework provided by care ethics, which emphasises the "moral salience of attending to and meeting the needs of the particular others for whom we take responsibility" (Held 2006). Just as the design of interfaces plays an important role, artists would have to support the notion of "slow care" by developing an increasingly critical stance towards excessive application of proprietary Creative-Ai, and an awareness of those being excluded from employing Creative-Ai in their creative processes.

An increased responsibility for artists, phrased as a positive motivation rather than a restricting norm, lies therefore at the base for developing a more sustainable creative-Ai. It is evident throughout the last decade that the funding of artists for collaborations with companies is increasing, both through public funding such as the European S+T+Arts[11] initiative and through funding at various IT companies for artist residencies. But what is the motivation for companies to fund such residencies? The presentation of artworks on company websites and the mentioning of a company as a sponsor in an artist portfolio are desirable publicity outcomes from the perspective of an IT company. In this way, the public image of products such as cloud computing is transformed from an abstract technological service to a mediator of a deeply human activity. Drott (2018) argues that music is foregrounded by companies providing streaming services to justify the particular value of the company platform. Whereas a focus on art dominates the public image of the companies, Drott (2018) demonstrates how essential parts of their business models are focused on advertisement and the marketing of user data. It remains to be discovered in how far companies will welcome initiatives by artists to reveal the environmental impact of projects conducted with their support.

In the context of political ecology, it is essential to critically analyse the principles of action and power structures that link the environmental impact of Creative-Ai with the cultural practices of artistic creation. As I have already elaborated, the means for computation are increasingly localised at a group of large companies, i.e. large capital functions at this point as a gatekeeper for artistic creation, and speculation may be a valuable means to investigate what might happen further down this road. Daughtry (2020) argues that empirical, quantitative approaches to the environmental impact of music fail to "get at the urgent ethical, political, and aesthetic questions that are tied up in music's relation to environment", and in this text I follow his suggestion that considers speculation as a suitable way to address these. My speculation is informed by a body of work that documents the role of the IT industry in reshaping the mediation of arts. In the following, I will focus on the example of music as the field of my expertise, but I consider this example to be strongly related to other forms of art. Hesmondhalgh & Meier (2018) document how the last 20 years have seen a shift from the consumer electronics to the IT industry, with the latter being now the primary sector that determines change in cultural technologies. This coincided with a strong re-structuring of the music industry towards streaming content and increased personalization. Notably, McGuigan & Manzerolle (2015) argue that the personalised mobile communication allows a more thoroughgoing commercialisation of culture than the previous mode of physical distribution, which was guided by inventions by the consumer electronics industry. This commercialisation of culture has been criticised for its potential to serve goals of user surveillance (Drott 2018), and the platforms that mediate between users and cultural content have been criticised for their opaqueness of user data handling actual and business goals (Eriksson et al 2019).

---

[11] https://starts.eu/

# Rethinking Attali

Allow me to leap backwards in time to a visionary text on the interactions between music and societal order, written well before the large impact of digitalisation that is the focus of the texts in the previous paragraph. Jacques Attali (1985) regarded music as a mirror of society and as a prophet for societal change. Hence, he argued that music brings into play certain forms of distribution networks, which reflect and even anticipate sociopolitical structures. The network that closely resembles the music industry in the era of the consumer electronics industry is what Attali calls the network of repetition. Here, recording enabled the stockpiling of music and a "cultural normalisation, and the disappearance of distinctive cultures." (Attali 1985, p. 111) Whereas Attali may be criticised for his strongly eurocentric perspective, this perspective provides an appropriate distortion that enables us to focus on the accumulation of capital in the Global North. In the context of the environmental impact of music, I follow Daughtry (2020) in his argument that such a distortion resulting from the focus on the economically dominating form is necessary to arrive at a conclusive overall picture. And in this distorted, simplified perspective I want to follow Attali in his suggestion of the network that he assumes will follow the network of repetition, which is the network of composition.

In this network, Attali foresaw that producing music would become an activity not undertaken for its exchange value, but only for the pleasure of the composers themselves. Composition would become a very localised activity, in which the distinction between consumption and production is resolved. For this to become true, appropriate instruments would need to be available that can be used by individuals to compose, leading to what could be termed as a democratisation of artistic creation. These creations would be made for the moment, not for stockpiling them in the shape of recordings, resulting in a permanently unstable and evolving practice. To contextualise this vision, Attali wrote this text in the 1970s, being a convinced advocate of socialism that he would later try to set in practice as advisor to the Mitterand government. But even back then, he saw risks in what may come when music is making a transition from the network of repetition to that of composition. He saw the risk of "emplacement of a new trap for music and its consumers, that of automanipulation" (Attali 1985, p.141), a risk increased by the fragility of meaning in the network of composition, and a risk that may be enlarged when the instruments provided for composition are themselves commodities controlled by large corporations.

The increasing fragmentation of music throughout the history of popular music into increasingly specialised genres (Bracket, 2016) has been regarded by Drott (2018) as culminating into disaggregating individual users. Along with such a process goes a destabilisation of meaning in cultural experiences caused by "radical and disorienting shifts" elicited by the constantly increasing push for innovation in the IT industry (Hesmondhalgh & Meier 2018). Several authors have seen a potential for democratisation and empowerment of artistic creation lying in the recent developments of creative-Ai (Mazzone & Elgammal 2019; Esling & Devis 2020). Hence, many harbingers indicate the approach of the network of composition, but who will provide the instruments? Here, Attali did not consider the composition conducted fully or partially by Ai-tools controlled by large companies. This scenario resembles the "musical holodeck" envisioned by Sturm et al (2019), a system that "provides any subscriber with limitless access to individualised musical experiences" and that makes any subscriber to such a system a composer. This, in fact, is Attali's network of composition with the means of production controlled by capital. Hence, in such a situation, the source of profit for a music industry is shifted from production and distribution of recordings to the provision of instruments for composition, turning the socialist utopia of Attali to the next stage of commodification of cultural activity by the IT industry.

## A case of commodification of artistic creation

Fortunately, the scenario elaborated in the previous section is purely fictional, and one may assume that even neoliberal venture capitalists are not adventurous enough to invest into a company that pursues such a goal. Wrong. The above-mentioned scenario seems to be well aligned with research and development at the largest music streaming provider, Spotify. The company's official presentation of their research (Spotify 2022) categorises research publications by Spotify employees into nine research areas, with music creation being one of them. However, among the 95 listed publications, only two have been associated with the research topic of music creation, whereas the most frequent topics are "Search and Recommendation" (43), "Machine Learning" (39), and "User Modelling" (28). Hence, the topics that R&D at Spotify focuses upon align well with the company's traditional core business of personalised music streaming. Research on the automatic creation of music, or even creative-Ai tools that support musicians in creative processes seem not to be part of the official Spotify R&D portfolio. In this context, it is remarkable that Spotify has founded the Creator Technology Research Lab (Spotify 2017), which explicitly focuses on the development of such tools. Publications from this lab seem to be excluded from the official research publication list of Spotify, exemplified by the absence of the publications of the head of the lab, Francois Pachet, from this list. Pachet is an internationally leading expert in Creative-Ai applied to music, and has recently co-published a book on the use of deep learning for the automatic generation of music (Briot et al 2020). More interestingly, his google scholar profile reveals his co-authorship in 11 US patent applications in 2020 and 2021 only. The abstracts of these patents, considered in relation to each other, create a picture of the compilation of tools needed for the automatic generation of music, such as the checking of an AI composition for plagiarism (Pachet & Roy 2022), a tool for building songs out of a collection of elementary stems (Pachet et al. 2022), and user evaluation tools to identify the best version of an automatic composition system (Pachet & Roy 2022b). Another part of the story (that Spotify is not telling) is their move to secure the rights to "create derivative works from User Content" of users of the Spotify for Artists app (Spotify 2021), which would include using user-uploaded music as training data for music generation systems[12]. Spotify for Artists was introduced in 2017, the same year in which the Creator Technology Research Lab was founded, and allowing derivative works from artists' uploads has been part of the terms and conditions since then until now. In June 2022, the business magazine Forbes (Hochberg 2022) reported Spotify's plans to publish a suite with tools that enable users to compose music with support of Ai, motivated by the idea that "users will engage more when they have a hand in creating the music with the help of AI".

Assessing the environmental impact of such a "listener/composer" system is impossible without detailed information of its components and its usage. Similar to the project by Nao Tokui and Qosmo introduced above, such a system is likely to be composed of several different deep learning (and other) architectures, an assumption supported by the diversity of architectures emphasised by Briot and Pachet (2020). Each of these needs to be trained, and - importantly - adapted and personalised to the needs of a listener/composer. This is the point where the transition from an individual artistic project to a commercial platform commodity becomes environmentally significant. In their first financial statement of 2022, Spotify declared to have "422 million monthly active users ("MAUs"), including 182 million Premium Subscribers". Even when we assume that only a minority of users would subscribe to a listener/composer service, and that personalisation will occur in clusters of users with similar demands, it is obvious that the carbon emission caused by a system like the one of Nao Tokui and Qosmo needs to be multiplied by some number between 1000 and 1 million.

---

[12] I was made aware of this aspect by Drott (2021).

Would such a situation be financially viable for the music content provider? At the current point, the profit of the streaming providers is limited by the large amount that they need to pay to music rights holders (Drott 2018). Compared to the outcry caused by Spotify populating playlists with fake artists (Goldschmitt 2020), the provision of an interactive listener/composer tool would likely be a more tolerated solution to increase profits from services. Given the current cost of a subscription at music streaming services, this money could be redirected from music rights holders to the energy expenditure for training systems. This approach is likely to be financially viable, but it would be connected to a large environmental cost. It would literally turn artist revenues into smoke.

At the current point of time, the share of carbon emissions assigned to streaming and AI development and application at music streaming providers is not known. The Equity and Impact Report by Spotify from 2021 (Spotify 2021b) acknowledges this, but provides little detail of how the reported emissions are related to specific causes within the business process. Whereas the reported net emission of Spotify (353054 MtCO2e) remains small in comparison to other IT companies such as Microsoft, the illustrated listener/composer system is likely to further increase the carbon footprint of streaming companies.

## Conclusion

The speculation that I presented in this chapter illustrates how the combination of the environmental impact of Creative-Ai with the increasing commodification of artistic creation by means of Creative-Ai is very likely to have severe consequences for cultural practices, the livelihood of artists, and the ecosystem. My examples focused on music, but a parallel example to the listener/composer system for music was presented in the form of DALL-E (OpenAI 2022). The model of access combines the creation of a number of free images with a paid model for additional images, indicating a similar business model as the one I assumed for the listener/composer service. Whereas the technical paper does not specify how much GPU-time it took to optimise the more than 3.5 billion parameters (Ramesh et al. 2022), the emissions of the whole development of this model are surely in the order of many road trips around the planet.

This chapter is one of the first attempts to promote awareness of the environmental impact of Creative-Ai. As I have explained, many open questions remain regarding how Creative-Ai is used in artistic processes. This needs to be addressed by conducting studies in collaboration with artists, for instance in the form of diary studies that document how exactly models are used. Such studies will then provide information on the diversity of the artistic processes, and the overall environmental impact as a consequence of model training and inference. A second aspect of our research is to develop new interfaces for artists that help to cultivate an attitude of slow care for the environment (Jääskeläinen et al 2022b). These studies aim at the development of tools that empower artists to monitor and visualize carbon impacts of their use of Ai models.

However, as we saw from the listener/composer example in this chapter, the environmental aspect is one perspective that calls for critique and action, whereas the cultural and social consequences of a commodification of automatic artistic creation are likely to be severe as well. A cross-sectional alliance between artists and academia is required to critically examine this overall impact of various Creative-Ai tools. In this context, it is important to reject technological determinism, and to emphasise the possibility of not employing Ai in certain application areas or promoting Ai infrastructures that are publicly accessible. Such a cross-sectional alliance must be sensitive to the social imbalances between Global South and North, as a demand for unconditional basic income for artists seems illusory in the context of the grim social conditions in most countries apart from the Northern centres of global capital.

Finally, another aspect of the political ecology of Creative-Ai is to establish a relation between the data used to train models, and the human beings that created that training data. New legal frameworks proposed by the European Union are likely to strengthen the legal position of such individual data creators (European Commission 2022), in case the proposals survive the lobbying by the large corporations who fear for their listener/composer and other models that would not be possible without free and exclusive access to huge amounts of user data. It is part of our future research to give a stronger voice to individuals and small enterprises in the process of forming data legislation. In this context, the proposition to consider data as labour instead of capital (Arrieta-Ibarra et al 2018) may indicate a possible way out of the upcoming crisis of artistic (and other) labour.